\documentclass[showpacs,aps,prl,twocolumn,amsmath,amssymb,superscriptaddress]{revtex4-1}
\usepackage{amsmath}           
\usepackage[latin1]{inputenc}  
\usepackage{graphicx}
\usepackage[dvipdf]{epsfig}
\usepackage{hyperref}
\usepackage{tabularx}

\begin{document}

\title{General approach for studying first-order phase transitions at 
low temperatures}

\author{C. E. Fiore}
\email{fiore@fisica.ufpr.br}
\author{M. G. E. da Luz}
\email{luz@fisica.ufpr.br}
\affiliation{\textit{Departamento de F\'isica,
    Universidade Federal do Paran\'a, 81531-980, Curitiba-PR, Brazil}}

\begin{abstract}
By combining different ideas, a general and efficient protocol to 
deal with discontinuous phase transitions at low temperatures is
proposed.
For small $T$'s, it is possible to derive a generic analytic expression 
for appropriate order parameters, whose coefficients are obtained 
from simple simulations.
Once in such regimes simulations by standard algorithms are not 
reliable, an enhanced tempering method, the parallel 
tempering -- accurate for small and intermediate system sizes 
with rather low computational cost -- is used.
Finally, from finite size analysis, one can obtain the
thermodynamic limit.
The procedure is illustrated for four distinct models, 
demonstrating its power, e.g., to locate coexistence lines and the 
phases density at the coexistence.
\end{abstract}


\maketitle

First-order phase transitions (FOPT) are ubiquitous in nature
\cite{books-first-order}, associated to a countless number of
processes \cite{lebowitz}.
Moreover, they take place in different temperature scales, e.g., 
from 2800 K in the Earth core-mantle \cite{oganov-matyska} to 
few and near zero Kelvin range (for many systems, in a rather
similar way \cite{kuwahara}), or even being responsible for 
unique effects, but across a very broad range of $T$'s \cite{silvera}.
In special, FOPT at low temperatures underpin important 
phenomena, like field-induced metal-insulator transitions, 
magnetoresistance, superfluidity, and Bose-Einstein condensation, 
among many others.

So, it is understandable and desirable the multitude of approaches 
(mainly numerically \cite{janke} given the difficulty to obtain 
general exact results \cite{pfister}) developed to study FOPT.
In certain instances, nevertheless, many of them can face procedural 
difficulties, not leading to precise results for the sough 
thermodynamical quantities, say, the exact location of coexistence 
lines. 

As particularly powerful methods we can cite cluster algorithms 
\cite{bouabci}, multicanonical \cite{berg} and Wang-Landau \cite{wang}.
In cluster, non-local configuration exchanges often ensure the crossing
of even high free-energy barriers.
But a drawback is its specialization: each model requires a specific 
and efficient algorithm to implement the transitions, not available 
in many cases.
On the other hand, Wang-Landau and multicanonical are general and
have been applied successfully in a great diversity of problems. 
However, the former may demand very large computational time 
to calculate the density of states, specially considering that
the number of states can increases very fast with the system 
size \cite{pla}.
The latter relies on histogram reweighting techniques to obtain
the appropriate averages, a difficult task for large systems
(see, e.g., \cite{okamoto}).

Given so, here we present a protocol to study FOPT at low 
temperatures by means of direct and simple numerical simulations. 
Extending previous results \cite{fiore-carneiro}, it considers 
around any transition a general parametric analytical expression 
for relevant thermodynamic quantities (like order parameter, density, 
magnetization, compressibility, etc).
The parameters are then obtained by simulating small systems, 
making the approach computationally fast.
Finally, from proper extrapolations, the correct thermodynamic 
limit is obtained.
Once standard algorithms usually fail for FOPT at low $T$'s,
even for small systems, we consider tempering methods (already 
shown reliable for FOPT, see \cite{fiore-luz-1,fiore-luz-2} 
and Refs. therein).
Thus, we use the parallel tempering, PT, very efficient for small and 
intermediate system sizes.
As we exemplify with four different lattices models, the approach
leads to a precise way to determine the coexistence regions.

We begin recalling a rigorous analysis for finite systems 
having $\mathcal{N}$ coexisting phases at $\xi^{*}$, with $\xi$ 
an appropriate phase transition parameter control (e.g.,
temperature or chemical potential).
It has been shown \cite{rBoKo}  that at low $T$'s and around 
$\xi^{*}$, the partition function is very accurately given by 
$Z = \sum_{n=1}^{\mathcal{N}} \, \alpha_n \, \exp[-\beta V f_n]$,
with $\beta = (k T)^{-1}$.
For the phase $n$, $f_n$ is the (metastable) free energy \cite{rBoKo}
per volume $V$ and $\alpha_n$ is the degeneracy, resulting from
eventual symmetries of the problem.

Next observe that $W = - \partial_\xi \ln[Z]/(\beta \, V)$ 
($\partial_x \equiv \partial /\partial x$) is frequently the start point to 
calculate distinct order parameters (density, magnetization, etc).
Since close to $\xi^{*}$, 
$f_n \approx f^{*} + f_n^{\prime *} \, y$ \cite{comment} for
$f_n^{\prime *} = \partial_{\xi^{*}} f_n(\xi^{*})$ and 
$y \equiv \xi - \xi^{*}$, we find the following general form for $W$:
\begin{equation}
W = (b_1 + \sum_{n=2}^{\mathcal{N}} b_n \, \exp[-a_n y])/
       (1 + \sum_{n=2}^{\mathcal{N}} c_n \, \exp[-a_n y]).
\label{eq1}
\end{equation}
The coefficients $a_n$, $b_n$ and $c_n$ depend on $\xi^{*}$, 
$f_n^{\prime *}$, $T$, and other system parameters.
But only the $a_n$'s are (linear) functions of $V$. 
Then, at the coexistence ($y=0$) $W$ is independent on 
the volume and for all $V$ the curves $W \times \xi$ cross at 
$\xi = \xi^{*}$.
In this way, Eq. (\ref{eq1}) can be used not only to locate the 
transition point, but also to determine the coexistence order 
parameters at the thermodynamic limit.
Moreover, if the $f$'s are ordered such that
$f_1^{\prime *} = f_2^{\prime *} = \ldots  = f_m^{\prime *} 
< f_{m+1}^{\prime *} < \ldots < f_{k}^{\prime *} = 
f_{k+1}^{\prime *} = \ldots = f_{\mathcal{N}}^{\prime *}$,
for $V\rightarrow \infty$ and $y \rightarrow 0^{\pm}$ we have 
$W_{\pm} = \sum_{n=v_{\pm}}^{n=u_{\pm}} b_n / c_n$, 
with $v_+ = 1, u_+ = m, v_- = k, u_- = {\mathcal N}$. 
For $k={\mathcal N}$ ($m=1$) $W_{+}$ ($W_{-}$) is given in terms 
of the sole phase which is immediately to the 
right (left) of $\xi^{*}$.

Thus, considering relatively small $V$'s we can obtain the 
parameters $a$'s $b$'s and $c$'s and hence, from the curves 
$W$, appropriate order parameters and response functions -- 
e.g., through derivatives of the order parameter -- at the 
transition point.
For instance, if $\xi = \mu$ is the chemical potential, 
$\rho(\mu,T) = - W$ is the density and 
$\chi = \partial_\mu \rho|_{T}$ is the isothermal compressibility.
Finally, from a simple scaling analysis \cite{w-janke}, but
using analytical smooth expressions, the thermodynamical properties 
are determined.

The above will work properly only with methods which 
correctly sample the configuration space \cite{bouabci,wang},
yielding reliable fittings for Eq. (\ref{eq1}) parameters.
Often, this is not so when systems displaying strong 
discontinuous transitions are simulated by conventional one-flip 
approaches, even for small sizes.
The solution is then to consider enhanced sampling, like parallel 
\cite{nemoto} and simulated \cite{marinari} tempering algorithms, 
PT and ST.
Since the former is particularly appropriate for FOPT (see 
\cite{fiore-luz-1} for details as well as for implementation), 
here we use the PT in our ``combo'' procedure for phase transitions 
at low $T$'s.

To illustrate the protocol, next we analyze four different lattice 
models displaying strong FOPT at low $T$'s.
In each case, what operationally sets a low temperature is the 
validity of the previous $Z$ decomposition.
Physically, it corresponds to $T$'s for which there is no overlap 
between the peaks of the order parameter bimodal distribution at 
the coexistence.
In all examples we perform accurate numerics with the PT 
algorithm and compare with the general Eq. (\ref{eq1}), whose 
parameters are always obtained using only four points 
from the simulations.

As the first example, we consider a rather complex system, the 
associative lattice-gas (ALG) model \cite{alg}, aimed to reproduce 
liquid polimorphism and water-like anomalies.
A site $i$ may or may not be occupied 
($\sigma_{\scriptscriptstyle{i}} =$ 1 or 0) by a molecule 
in a triangular lattice.
The orientational variable $\tau_{\scriptscriptstyle{i}}^{ij} = 0, \, 1$
represents the possibility of hydrogen bonding (in a maximum 
of four) between the molecule in site $i$ and those in the adjacent six 
neighbors $j$, provided $\tau_{i}^{ij} \tau_{j}^{ji} = 1$.
Two first neighbor molecules have an interaction energy of $-v$
($-v + 2 u$) if there is (there is not) a hydrogen bond between them.
The Hamiltonian reads
\begin{equation}
{\cal H} = 2 u \sum_{<i,j>} \sigma_{i} \sigma_{j} [(1 - v/(2 u)) -  
\tau_{i}^{ij} \tau_{j}^{ji}] - \mu \sum_{i}\sigma_i.
\end{equation}
It presents one gas and two liquid, LDL and HDL, phases
of densities $\rho = 0$, $\rho=3/4$, and $\rho=1$, respectively.
For fixed $T$, by increasing $\mu$ we pass through two FOPT, 
namely, gas-LDL and LDL-HDL.

\begin{figure}[top]
\centerline{\psfig{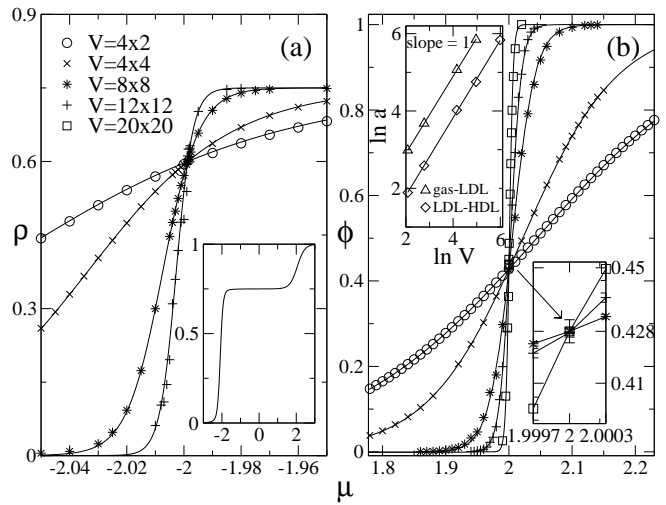}} 
\caption{The ALG model for the parameters as in the text and
different $V = L \times L$.
The continuous lines are curves from Eq. (\ref{eq1}) for:
(a) $\rho \times \mu$ around the gas-LDL transition (the inset 
shows $\rho \times \mu$ in the whole $\mu$ range where the two 
FOPS take place); and
(b) $\phi (= 4 \rho - 3) \times \mu$ around the LDL-HDL transition.
The blow up shows the crossing for $L=8, 12, 20$.
The expected linear $a \times V$ behavior is displayed 
in the inset of (b) (the log-log scale is just for accommodating
both cases).
For comparison, the simulations for $V=4\times 2$ are exact.}
\label{fig1}
\end{figure}

We study the ALG model gas-LDL and LDL-HDL FOPT by plotting 
the density vs $\mu$ for $T=0.300$, $u=v=1$, and different 
$V$'s.
For the gas-LDL case, we show the results in Fig. \ref{fig1} (a).
We clearly see a good coincident crossing of all curves at 
$\mu = -1.9986(2)$, for $\rho = 0.600(1)$.
The exact transition density $\rho = 3/5 = 0.6$ is understood 
recalling that at the coexistence both gas ($\rho=0$) and LDL 
($\rho = 3/4$) phases have equal weight. 
Given that $\alpha_{LDL} = 4$, the value follows.
Around LDL-HDL, $\rho$ does not vanish, inset of Fig. \ref{fig1} (a).
Since $3/4$ (the totality) of the lattice is filled by molecules 
in the LDL (HDL) phase, a better order parameter is the rescaled
density $\phi = (4 \rho - 3)$.
Thus, in Fig. \ref{fig1} (b) we display $\phi \times \mu$
for the LDL-HDL transition. 
Again, all the isotherms are well described by Eq. (\ref{eq1}), 
crossing at $\mu = 2.0000(3)$ with $\rho = 0.857(1)$.
In Fig. \ref{fig1} (b) inset we confirm the expected 
linear dependence on $V$ for the parameter $a_2 = a$ 
(a single $a$ once we have only two phases in each transition).

Next we consider the Bell-Lavis (BL) model, which also displays
water-like anomalies.
The sites may or may not be occupied ($\sigma_{\scriptscriptstyle{i}}=$ 
1 or 0) by molecules of two possible orientations.
But differently from the ALG, the van-der-Waals interaction between 
two adjacent molecules is attractive, $-\epsilon_{vdw}$.
So, there is no energetic punishment if hydrogen bonds (of
energy $-\epsilon_{hb}$) are not formed.
Such distinctions from the ALG, e.g., result in a second order phase
transition for LDL-HDL, but still a FOPS for the gas-LDL.
It is described by ($\zeta \equiv \epsilon_{vdw}/\epsilon_{hb}$)
\begin{equation}
{\mathcal H} = - \epsilon_{hb} \, \sum_{<i,j>} \sigma_{i} \, \sigma_{j} \,
[\tau_{i}^{ij} \, \tau_{j}^{ji} + \zeta] - \mu \sum_{i} \sigma_{i}.
\label{hambl}
\end{equation}
For $\zeta <1/3$, the BL presents three phases, gas 
($\rho = 0$), LDL ($\rho = 2/3$ in a honeycomb structure),
and HDL ($\rho = 1$) \cite{bell}.
In the numerics we set $\epsilon_{hb} = 1$ and $\epsilon_{vdw} = 1/10$.  

\begin{figure}[top]
\centerline{\psfig{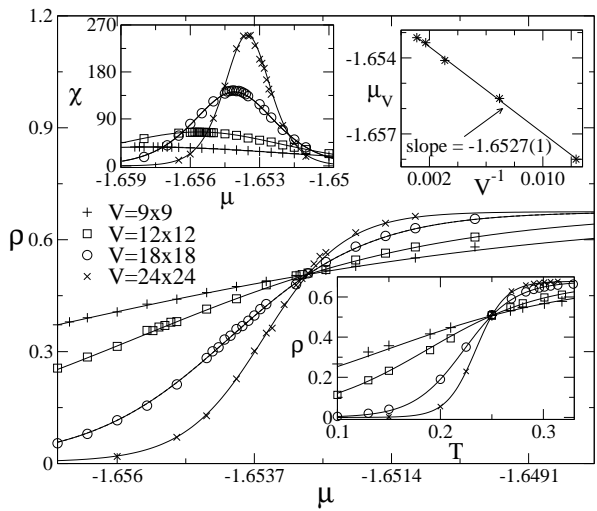}} 
\caption{For the BL model gas-LDL FOPT and parameters as
in the text, $\rho \times \mu$ for distinct $V$'s and $T=0.25$. 
All the continuous lines are properly obtained from Eq. (\ref{eq1}). 
The upper insets show the isothermal compressibility $\chi$ 
vs $\mu$ and $ \mu_V$ (the values of $\mu$ at the peaks of $\chi$)
vs $V^{-1}$.
The lower inset shows $\rho \times T$ curves for $\mu = -1.6528$,
which cross at $T=0.25$.} 
\label{fig2}
\end{figure}

For $T=0.25$, in Fig. \ref{fig2} we plot $\rho \times \mu$ around 
the FOPT gas-LDL.
Once more, the simulations are well described by Eq. (\ref{eq1}). 
The isotherms cross at $\mu =-1.6528(1)$, with 
$\rho \approx 0.507(2)$ very close to the exact $\rho=1/2$
(which can be inferred as done for the ALG model).
In the upper-left inset we show
$\chi = (\frac{\partial \rho}{\partial \mu})_{T}$ by 
properly differentiating Eq. (\ref{eq1}) (continuous lines)
and by numerically simulating  
$\chi = V (\langle \rho^2\rangle-\langle \rho \rangle^{2})$. 
Note the remarkable agreement, again illustrating the power of
Eq. (\ref{eq1}) to describe relevant thermodynamic quantities 
around FOPT.
The upper-right inset displays the values of $\mu = \mu_V$ (for 
which $\chi$ is maximum) vs $V^{-1}$.
This type of scaling extrapolation also can give the thermodynamic 
limit for the transition, here $\mu=-1.6527$, basically the 
same value obtained from the crossing.
Finally, instead of $\mu$ one could take $T$ as the control 
parameter.
Setting $\mu =-1.6528$ and varying $T$ we see in the lower 
inset of Fig. \ref{fig2} the gas-LDL transition.
As it should be, the curves cross at $T=0.25$, with $\rho \approx 1/2$.
Finally, we note that for $T > 0.43$, the results from the present
method starts to be less accurate.

The Blume-Emery-Griffiths (BEG) model yields
\cite{BEGMODEL}
\begin{equation}
 {\cal H} = - \sum_{<i,j>} [J \sigma_{i} \, \sigma_{j} 
            + K \sigma_{i}^{2}\sigma_{j}^{2}] 
            - \sum_{i}  [H \sigma_i - D \sigma_i^2],
\label{e3}
\end{equation}
where a site $i$ is either empty or occupied by two different 
type of species ($\sigma_i = 0, \pm 1$).
Parameters $J$ and $K$ are interaction energies and $D$ and $H$ 
denote linear combination of the species $\mu$'s.
This system is a particularly interesting test because the otherwise 
very reliable cluster algorithm for the BEG model \cite{bouabci} 
fails for some particular $K/J$'s,
e.g., the value we address $K/J=-0.5$. 
So, for a better comparison with our procedure, we also
propose a new cluster-Metropolis hybrid approach, which includes 
intermediary Metropolis algorithm steps (details will appear 
elsewhere).
We note, nevertheless, that the Metropolis alone is not able to cross 
the high free energy barriers at the phases coexistence.

\begin{figure}[top]
\centerline{\psfig{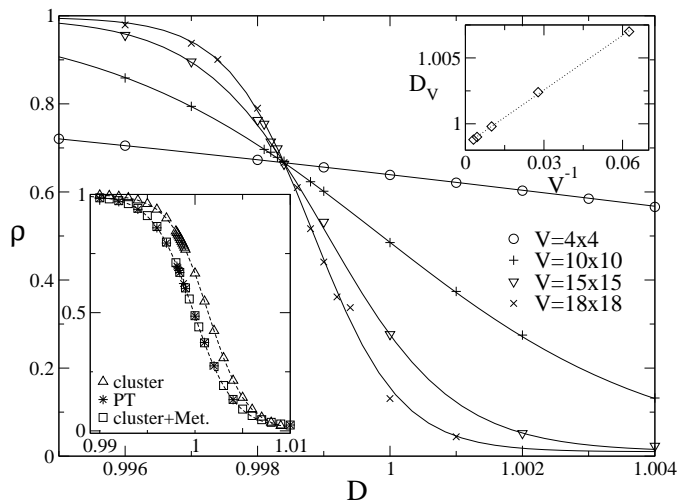}} 
\caption{For the BEG model, $\rho \times \mu$ for parameters
as in the text and distinct $V$'s. 
The results for $V=4 \times 4$ are exact.
The continuous lines properly come from Eq. (\ref{eq1}). 
The right inset shows $D_V$ ($D$ for which $\chi$ is maximum)
vs $V^{-1}$.
The left inset compares numerical simulations from the PT,
cluster and hybrid cluster-Metropolis algorithms for 
$V = 10 \times 10$.}
\label{fig3}
\end{figure}

In Fig. \ref{fig3} we plot $\rho \times D$ for $H=0$, $J=1$, 
$K=-0.5$, $T=0.20$, and different $V$'s.
We have a FOPT with all the isotherms crossing at $D = 0.9984(1)$ 
and $\rho \approx  2/3$. 
The right inset of Fig. \ref{fig3} shows the position $D_V$
of the peak of $\chi = -(\frac{\partial \rho}{\partial D})_{T} $, 
calculated directly from Eq. (\ref{eq1}). 
A linear extrapolation of $D_V \times V^{-1}$ gives $D=0.99845(5)$, 
in excellent agreement with the crossing value.
For $V = 10 \times 10$, we plot in the left
inset simulations from the usual cluster, the improved (but 
dedicated) cluster-Metropolis, and PT algorithms.
The latter two display very good concordance, with the cluster 
given poorer results.
We should mention that for the BEG and ALG models there are 
no precise simulations in the literature for the parameter
conditions here considered.

Lastly, we discuss the asymmetric Ising Hamiltonian on a triangular 
lattice (of sublattices $(\alpha,\beta,\gamma)$) \cite{landau86}
\begin{equation}
{\cal H} = -J \sum_{<i,j>}\sigma_{i}\sigma_{j} - K
\sum_{<i,j,k>}\sigma_{i}\sigma_{j}\sigma_{k} - H \sum_{i}\sigma_{i}.
\label{ising3nn} 
\end{equation}
The second sum is over first neighbors trios forming 
triangles. 
Using the Wang-Landau method \cite{wang}, the model has been studied 
in details \cite{wang2} (but for larger systems and lower numerical 
precision). 
It displays one ferrimagnetic, $(--+)$, and two ferromagnetic,
$(+++)$ and $(---)$, phases. 
For very low temperatures, by increasing $H$ the system displays a 
second-order phase transition $(---) \rightarrow (--+)$ and then a 
FOPT $(--+) \rightarrow (+++)$. 
The $(--+)$ phase disappears in a critical endpoint ($T_c = 2.443(1), 
H_c = -2.934(1)$), above it giving rise only to a FOPT between the 
two ferromagnetic phases. 
Although the magnetization per site $m$ is not the actual order 
parameter, for rather small system sizes we can extract from it
any relevant FOPT information.

\begin{figure}[top]
\centerline{\psfig{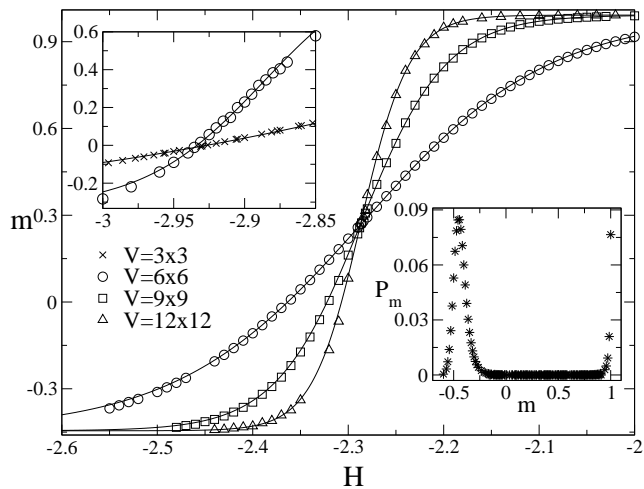}} 
\caption{For the asymmetric Ising model and parameters as in the
text, $m \times H$ for distinct $V$'s and $T=5.00$ ($T=2.40$, left
inset). 
Continuous lines are the curves from Eq. (\ref{eq1}).
The right inset shows $P_m \times m$ for the $T=5.00$ case.}
\label{fig4}
\end{figure}

In Fig. \ref{fig4} we plot $m \times H$ for $J=1, K=2$ and $T=5.00$
for the $(---) \rightarrow (+++)$ FOPT.
We see that Eq. (\ref{eq1}) represents quite well the transition.
In the right inset we show the histogram magnetization 
density probability 
$P_m$ vs $m$ for $L=12$, $T=5.00$ and $H=-2.3325$, 
illustrating further that the phases coexistence is being 
properly characterized \cite{fiore-luz-1}.
Likewise, the FOPT  $(--+) \rightarrow (+++)$ for $T=2.40$ in
the left inset is well described by our method.
All the isotherms cross at $H=-2.2863(5)$ and $H=-2.9357(5)$
(left inset), in fair agreement (given the different numerical
accuracies) with the estimates $H=-2.284(1)$ and $H=-2.939(1)$ 
by the authors of Ref. \cite{wang2} (private communication).

By considering Eq. (\ref{eq1}), derived from rigorous results
at low $T$'s, we have proposed a general protocol to study FOPT.
It is accurate and demands only few simulations for 
relatively small systems, hence a computationally low 
cost procedure.
The approach has been very successfully applied to four distinct 
lattice models.
Of course, more analyzes, e.g., for higher dimensions and continuous 
systems (presently under progress, but with promising preliminary 
findings) are in order as further tests.
Nevertheless, we believe the method already shows itself a valuable 
tool to analyze the very important problem of FOPT at low temperatures.

Research grants are provided by CNPq and CT-Infra.

\end{document}